\documentclass[compsoc,conference,a4paper,10pt,times]{IEEEtran}
\usepackage{cite}
\usepackage{amsmath,amssymb,amsfonts}
\usepackage{algorithmic}
\usepackage{graphicx}
\usepackage{textcomp}
\usepackage{xcolor}
\usepackage{subcaption}

\begin{document}
\bstctlcite{IEEEexample:BSTcontrol}

\title{Poster: Towards an Automated  Security Testing Framework for Industrial UEs}

\author{
{\rm Sotiris Michaelides$^{\ast}$, Daniel Eguiguren Chavez$^{\ast}$, and Martin Henze$^{\ast\mathsection}$} \\
    {\rm $^\ast$\textit{Security and Privacy in Industrial Cooperation}, RWTH Aachen University, Germany} \\
    {\rm $^\mathsection$\textit{Cyber Analysis \& Defense}, Fraunhofer FKIE, Germany} \\
    {\rm michaelides@spice.rwth-aachen.de $\cdot$ daniel.eguiguren@rwth-aachen.de $\cdot$ henze@spice.rwth-aachen.de}
}

\maketitle

\begin{abstract}
With the ongoing adoption of 5G for communication in industrial systems and critical infrastructure, the security of industrial UEs such as 5G-enabled industrial robots becomes an increasingly important topic.
Most notably, to meet the stringent security requirements of industrial deployments, industrial UEs not only have to fully comply with the 5G specifications but also implement and use correctly secure communication protocols such as TLS.
To ensure the security of industrial UEs, operators of industrial 5G networks rely on security testing before deploying new devices to their production networks.
However, currently only isolated tests for individual security aspects of industrial UEs exist, severely hindering comprehensive testing.
In this paper, we report on our ongoing efforts to alleviate this situation by creating an automated security testing framework for industrial UEs to comprehensively evaluate their security posture before deployment.
With this framework, we aim to provide stakeholders with a fully automated-method to verify that higher-layer security protocols are correctly implemented, while simultaneously ensuring that the UE's protocol stack adheres to 3GPP specifications.
\end{abstract}


\section{Introduction}

5G not only represents the latest generation of mobile networks, serving billions of users globally~\cite{koutsos20195g}, but also is the first mobile technology to provide low latency and high reliability guarantees required for its utilization in industrial systems and critical infrastructure \cite{MICHAELIDES2025107645}.

However, the integration of 5G into industrial environments substantially expands the attack surface of security-critical environments~\cite{vidal2017poster,bodenhausen2023challenges}, as new components and communication interfaces are introduced into the system architecture~\cite{henze2024sierra}. 
Although 5G introduces enhanced security mechanisms compared to previous generations, many of these improvements are  optional features, leaving network operators with the discretion to implement them or not \cite{MICHAELIDES2025107645}.
Experience from real-world deployments shows that security features are either not being implemented~\cite{lasierra2023european} or not configured/implemented correctly~\cite{251578}.

Likewise, the increasing interconnection of industrial systems  and resulting broadened attack vectors \cite{bader2023wattson} require the use of secure communication protocols, e.g., TLS~\cite{10.1145/3488932.3497762}.
When integrating 5G into industrial networks, their use becomes imperative as 
(i) user plane (UP) protection in 5G does not provide end-to-end security and
(ii) optional UP protection is often disabled due to its substantial impact on network latency \cite{bicmac,tlscore,michaelides2025latency}, which threatens real-time data transmission as required by industrial use cases.
However, large-scale measurements of industrial deployments show that security protocols are often either not used at all or configured insecurely~\cite{10.1145/3488932.3497762,dahlmanns2020easing}.

Consequently, before deploying industrial user equipments (UEs) such as 5G-enabled industrial robots, operators of industrial networks should first check whether both the 5G functionality and security protocols used for industrial communication are implemented and configured securely.
While both scientific approaches and practical tools exist to perform such tests \emph{for individual aspects}~\cite{iker,tlsfuzzer,testsslsh,reloaded,astra}, a comprehensive framework that allows to automate the execution and evaluation of security tests for both the 5G part and the secure implementation of industrial communication on top of 5G are missing.

\textbf{Our vision.}
To support the \emph{secure} integration of 5G-enabled industrial devices, we envision an \emph{automated} security testing framework that assists operators to \emph{comprehensively} test industrial UEs for the correct implementation and configuration of all relevant security features.
Such a framework should not only cover 5G specific aspects such as a device complying to unauthenticated commands~\cite{reloaded}, but also ensure that upper layer security protocols such as TLS are used and correctly configured, e.g., not using outdated ciphers.
In the end, such a framework should empower operators to verify the security posture of safety-critical industrial components \emph{before} they get deployed to a production 5G environment.

\textbf{Our contributions.} 
In this paper, we report on our ongoing work in developing the first automated security framework specifically tailored for examining the security posture of industrial UEs.
To this end, we build upon previous research to integrate existing and novel security tests into a unified framework that automatically tests the compliance of the industrial UE's 5G protocol stack with the 3GPP specifications as well as the correct implementation and configuration of secure industrial protocols, especially those relying on TLS.

\section{Technical Background \& Related Work}

To lay the foundation for our work, we introduce the most important aspects of 5G and discuss related work.

\begin{figure}[t]
    \centering
    \includegraphics[width=\linewidth]{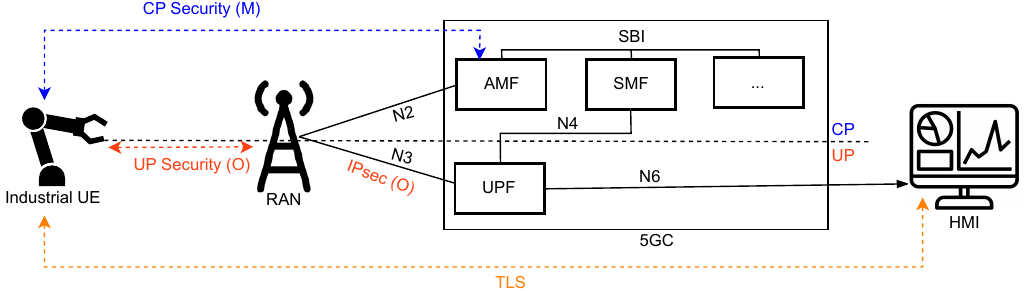}
    \caption{High-level architecture of an industrial 5G network showing the industrial UE, RAN, and 5GC with associated planes and security controls. }

    \label{fig:5g_architecture}
\end{figure}

\textbf{5G Components.} As depicted in Fig.~\ref{fig:5g_architecture}, an industrial 5G network consists of three primary components that together manage both \emph{Control Plane (CP)} and \emph{User Plane(UP)} traffic: the \emph{industrial User Equipment (UE)}, the \emph{Radio Access Network (RAN)}, and the \emph{5G Core (5GC)}.
The UE is the end-user device (such as industrial sensors or robots) paired with authentication credentials. 
UEs initiate connections and generate both UP data (e.g., application traffic) and CP signaling (e.g., session setup, mobility management). 
To facilitate this, UEs employ the \emph{Radio Resource Control (RRC)} protocol~\cite{rrc} to communicate with the RAN and the \emph{Non-Access Stratum (NAS)} protocol~\cite{nas} to interact with the 5GC.
The RAN serves as the intermediary between the UE and the 5GC, handling all wireless communication. 
Its responsibilities include among others, managing radio resources and enforcing Quality of Service. 
The RAN also plays a dual role by relaying both CP signaling (i.e., NAS ) and UP data between the UE and the 5GC.
At the core of the network, the 5GC consists of modular network functions responsible for processing and routing both types of traffic. 
CP functions include the Access and Mobility Management Function (AMF), which handles tasks such as authentication and mobility management. 
On the UP, the User Plane Function (UPF) is responsible for packet forwarding and routing user data to external data networks.

\textbf{5G Planes \& Security.} \emph{CP} data, exchanged between network components, is mandated by the 5G specification to be end-to-end integrity protected between the UE and the 5GC (specifically, the AMF). These signaling messages are critical for reliable network operation and support key functions such as session management and mobility handling.
In contrast, \emph{UP} data—the actual application (e.g., industrial) data transmitted by the end device—is only optionally protected. Even when optional security mechanisms are enabled (cf. Fig.~\ref{fig:5g_architecture}), they do not ensure true end-to-end protection.
Interfaces such as N6, which connects the UPF to external data networks, lack standardized security and are left to operator discretion.
These gaps highlight the necessity of higher-layer security protocols (e.g., TLS) to ensure  end-to-end protection of user data—especially in latency-sensitive deployments.

\textbf{Related Work.} 
Prior works focus on testing either TLS or 5G CP protocols.
Our work bridges these approaches by proposing the first unified framework integrating tests from both fields and  automating testing.

Several tools test the security of implementations of security protocols such as TLS and IPsec (e.g., \cite{iker,tlsfuzzer}), mostly offering similar functionality.
In this work, we use \emph{testssl} \cite{testsslsh}, a lightweight yet comprehensive command-line tool for automated testing of TLS security features on server endpoints.
It provides detailed reports on supported versions and cipher suites, analyzes preconfigured settings for vulnerabilities or misconfigurations, and detects known TLS exploits such as Heartbleed and LOGJAM. 

To assess the correctness of individual aspects of the 5G control plane, several academic testing frameworks have been developed. Bitsikas et al.~\cite{reloaded} introduced a tool to evaluate UE compliance with 3GPP NAS and RRC specifications, using modified Open5GS and srsRAN instances along with JSON-defined test cases. Upon receiving specific uplink messages, the framework hijacks the 5GC or RAN to inject custom downlink commands, logging interactions via PCAPs and system logs.
Building on this, Khandker et al.~\cite{astra} proposed an automated framework for NAS-layer testing. It generates test cases from user input, executes them via modified network components, and evaluates UE responses using rule-based or LLM-based analysis against 3GPP specifications.

\section{Automated Security Testing Framework}

While prior work tests individual security aspects of industrial UEs, we strive to integrate a comprehensive set of tests into a unified and automated framework.

\textbf{Concept.}
Our framework assesses the security posture of industrial UEs by testing the specification compliance of their 5G protocol stack implementation as well as the correct implementation and configuration of upper layer security  protocols.
We focus on widely adopted protocols that are secure when properly configured.
Our tests aim to identify known vulnerabilities (e.g., TLS Heartbleed) or insecure configurations that could weaken security.
For instance, using deprecated algorithms such as RC4 in TLS can expose communications to known attacks.

\begin{figure}[t]
    \centering
    \includegraphics[width=\linewidth]{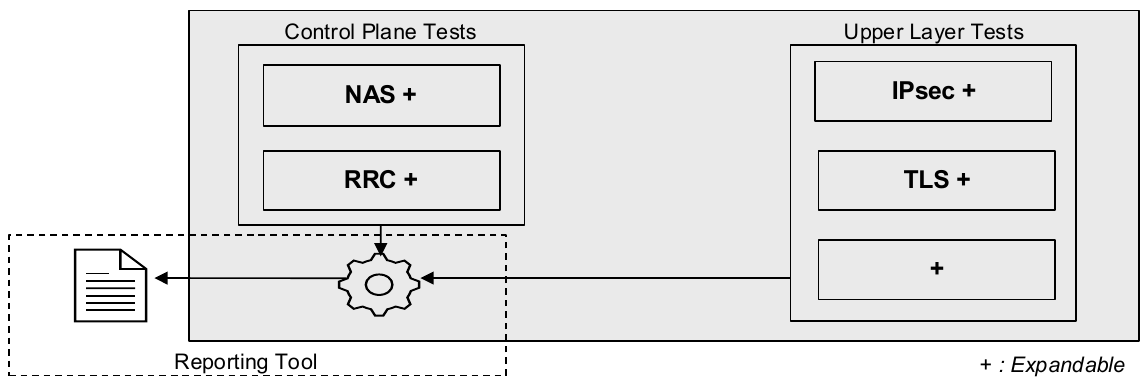}
    \caption{Overview of our framework concept. Its modular design enables the integration of additional test cases and new upper layer protocols.}
    \label{fig:concept}
\end{figure}

As depicted in Fig.~\ref{fig:concept}, our automated security test framework integrates tests for the correct implementation of the control plane primary signaling protocols: \emph{RRC} and \emph{NAS} as well as for the correct implementation and configuration of upper layer security protocols such as \emph{TLS} and \emph{IPsec}.
Our framework not only automates the execution of such tests but further more unifies the representation of reported test results.

\textbf{Implementation.}
The novelty of our approach lies in the full automation of multiple security tests, which is crucial for industrial UEs that rely on secure communication protocols. 
To showcase the feasibility of our approach, we combine and extend three existing testing tools.

\begin{figure}[t]
    \centering
    \includegraphics[width=\linewidth]{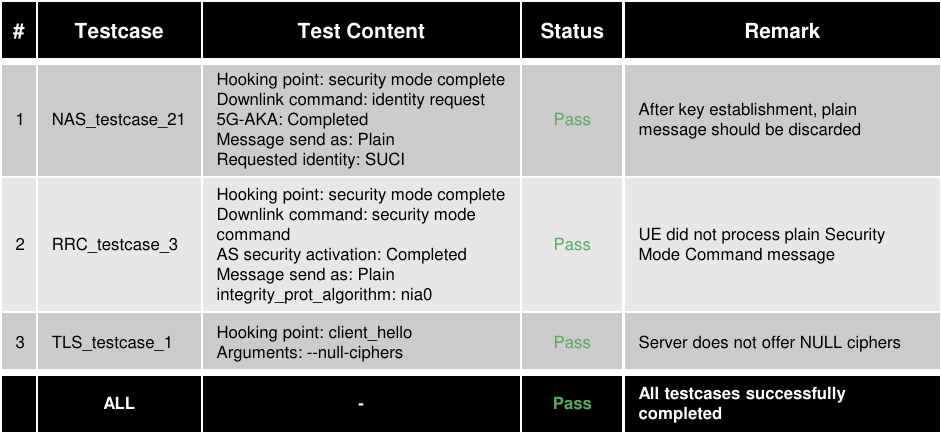}
    \caption{Our framework consolidates all security testing results in one unified, reader-friendly report.}
    \label{fig:report}
\end{figure}

The first component is the tool developed by Bitsikas et al.~\cite{reloaded}, which conducts a series of static, predefined test cases on the CP of 5G, covering both the NAS and RRC layers. 
The second component is \emph{testssl}~\cite{testsslsh}, which performs a series of tests on the TLS protocol. The third component is based on the evaluation tool by Khandker et al.~\cite{astra}. 
We have expanded its evaluation capabilities for a defined subset of RRC-layer test cases, enabling the framework to examine the UE's behavior against the 3GPP specification~\cite{nas, rrc} on both the NAS and RRC layers. Additionally, we added support for the automatic verification of TLS  test case results against the BSI recommendations.
Our framework consolidates the results of these different tests to generate a unified, reader-friendly report as exemplarily depicted in Fig.~\ref{fig:report}.

\section{Preliminary Results}
\begin{figure}[t]
    \centering
    \includegraphics[width=\linewidth]{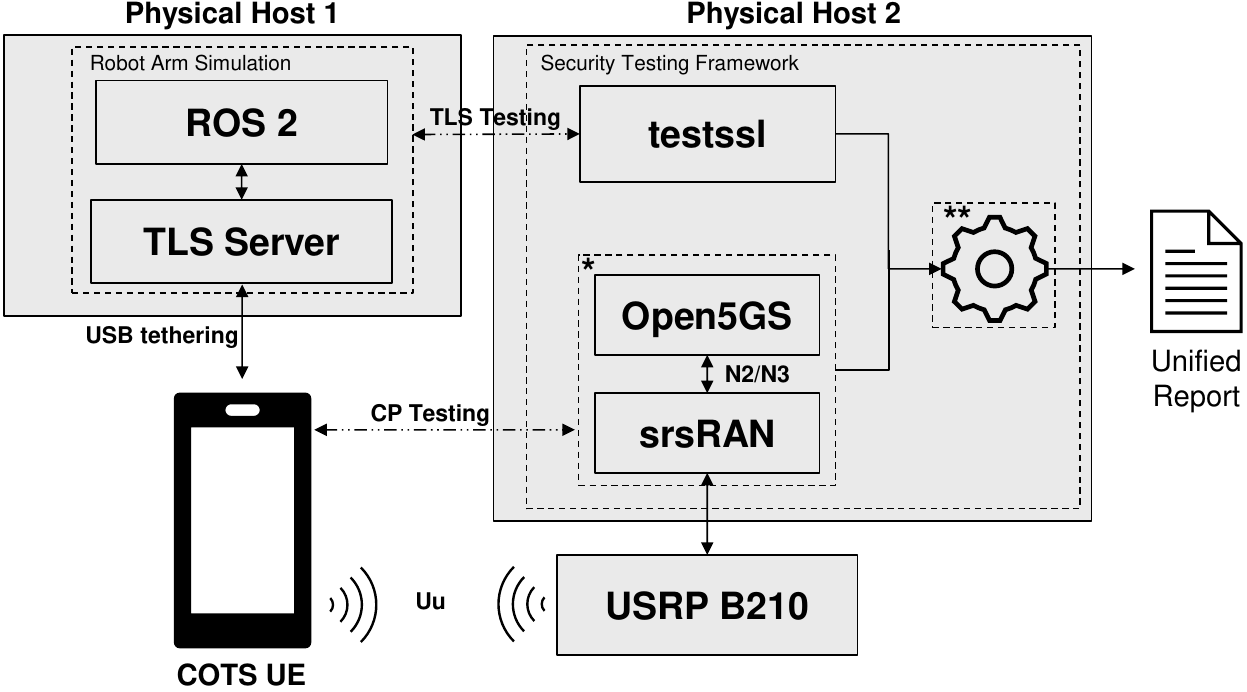}
    \footnotesize \textit{~*\cite{reloaded}, **\cite{astra} expanded with RRC evaluation}
    \caption{Our setup used to gather preliminary results.}
    \label{fig:implementation}
\end{figure}

As a proof of concept, we deploy our framework in a realistic setup as shown in Fig.~\ref{fig:implementation} to perform preliminary testing.
We utilize ROS2, which provides a robotic arm simulation, with all traffic sent by the robot secured using TLS. 
We further employ a UE with a MediaTek Dimensity 700 baseband as a gateway for the robot, handling all 5G communication.
The 5G network is realized using Open5GS, srsRAN, and a URSP B210.
We run our automated testing framework on a machine equipped with 128 GB of RAM and an Intel Core i9-14900 processor.

\textbf{Testing Performance.} In its current state, our framework implements security testing for TLS using the \textit{testssl} tool, supporting all of its available features. 
It also includes 53 test cases for the NAS layer and 16 for the RRC layer.
In terms of performance, all tests are executed in approximately 30 minutes. 
Of this, \(\sim \)99\% is dedicated to the CP tests, as each test requires a full network restart.

\textbf{Initial Findings.} While our current focus is on expanding test case coverage for 5G signaling protocols and enhancing unified reporting, preliminary testing using our real-world setup and the expanded RRC evaluation has already uncovered a vulnerability in the RRC layer of a tested UE which may lead of the extraction of UE capabilities before security activation.
We are in the process of disclosing this vulnerability to the manufacturer.

\section{Conclusion \& Future Work}

With our automated security testing framework, we aim to provide industrial companies with a fully automated tool to efficiently test industrial UEs before deploying them into the production network. Our prototype implementation, which has already been deployed and tested, uncovered a vulnerability in the MediaTek Dimensity 700.

Future work will focus on expanding the CP tests as well as adding more upper layer security protocols, e.g., IPsec, and secure industrial protocols, e.g., OPC UA. 
Subsequently, we will leverage our framework to comprehensively test commercial industrial UEs.

\vspace{1em}
\textbf{Acknowledgments.}
Funded by the German Federal Office for Information Security (BSI) under project funding reference number 01MO24003B (CSII). The authors are responsible for the content of this publication.

\bibliographystyle{IEEEtran} %

\begin{thebibliography}{10}
\providecommand{\url}[1]{#1}
\csname url@samestyle\endcsname
\providecommand{\newblock}{\relax}
\providecommand{\bibinfo}[2]{#2}
\providecommand{\BIBentrySTDinterwordspacing}{\spaceskip=0pt\relax}
\providecommand{\BIBentryALTinterwordstretchfactor}{4}
\providecommand{\BIBentryALTinterwordspacing}{\spaceskip=\fontdimen2\font plus
\BIBentryALTinterwordstretchfactor\fontdimen3\font minus \fontdimen4\font\relax}
\providecommand{\BIBforeignlanguage}[2]{{%
\expandafter\ifx\csname l@#1\endcsname\relax
\typeout{** WARNING: IEEEtran.bst: No hyphenation pattern has been}%
\typeout{** loaded for the language `#1'. Using the pattern for}%
\typeout{** the default language instead.}%
\else
\language=\csname l@#1\endcsname
\fi
#2}}
\providecommand{\BIBdecl}{\relax}
\BIBdecl

\bibitem{koutsos20195g}
A.~Koutsos, ``{The 5G-AKA authentication protocol privacy},'' in \emph{IEEE EuroS\&P}, 2019.

\bibitem{MICHAELIDES2025107645}
S.~Michaelides \emph{et~al.}, ``{Secure Integration of 5G in Industrial Networks: State of the Art, Challenges and Opportunities},'' \emph{Future Generation Computer Systems}, 2025.

\bibitem{vidal2017poster}
J.~M. Vidal \emph{et~al.}, ``{Poster: Mitigation of DDoS attacks in 5G networks: A bio-inspired approach},'' in \emph{IEEE EuroS\&P}, 2017.

\bibitem{bodenhausen2023challenges}
J.~Bodenhausen \emph{et~al.}, ``{Securing Wireless Communication in Critical Infrastructure: Challenges and Opportunities},'' in \emph{MobiQuitous}, 2023.

\bibitem{henze2024sierra}
M.~Henze \emph{et~al.}, ``{Towards Secure 5G Infrastructures for Production Systems},'' in \emph{ACNS}, 2024.

\bibitem{lasierra2023european}
O.~Lasierra \emph{et~al.}, ``{European 5G security in the wild: Reality versus expectations},'' in \emph{ACM WiSec}, 2023.

\bibitem{251578}
D.~Rupprecht \emph{et~al.}, ``Call me maybe: Eavesdropping encrypted {LTE} calls with {ReVoLTE},'' in \emph{USENIX Security}, 2020.

\bibitem{bader2023wattson}
L.~Bader \emph{et~al.}, ``{Comprehensively Analyzing the Impact of Cyberattacks on Power Grids},'' in \emph{IEEE EuroS\&P}, 2023.

\bibitem{10.1145/3488932.3497762}
M.~Dahlmanns \emph{et~al.}, ``{Missed Opportunities: Measuring the Untapped TLS Support in the Industrial Internet of Things},'' in \emph{ACM ASIA CCS}, 2022.

\bibitem{bicmac}
T.~Heijligenberg \emph{et~al.}, ``{BigMac: Performance Overhead of User Plane Integrity Protection in 5G Networks},'' in \emph{ACM WiSec}, 2023.

\bibitem{tlscore}
O.~Zeidler \emph{et~al.}, ``{Performance Evaluation of Transport Layer Security in the 5G Core Control Plane},'' in \emph{ACM WiSec}, 2024.

\bibitem{michaelides2025latency}
S.~Michaelides \emph{et~al.}, ``{Assessing the Latency of Network Layer Security in 5G Networks},'' in \emph{ACM WiSec}, 2025.

\bibitem{dahlmanns2020easing}
M.~Dahlmanns \emph{et~al.}, ``{Easing the Conscience with OPC UA: An Internet-Wide Study on Insecure Deployments},'' in \emph{ACM IMC}, 2020.

\bibitem{iker}
Borja \emph{et~al.}, ``Iker,'' \url{https://github.com/libcrack/iker}, 2015.

\bibitem{tlsfuzzer}
A.~Kario, ``Tlsfuzzer: Ssl and tls protocol test suite and fuzzer,'' \url{https://github.com/tlsfuzzer/tlsfuzzer}, 2025.

\bibitem{testsslsh}
D.~Wetter and contributors, ``testssl.sh: A free command line tool for testing tls/ssl encryption,'' \url{https://testssl.sh/}, 2025.

\bibitem{reloaded}
E.~Bitsikas \emph{et~al.}, ``{UE Security Reloaded: Developing a 5G Standalone User-Side Security Testing Framework},'' in \emph{ACM WiSec}, 2023.

\bibitem{astra}
S.~Khandker \emph{et~al.}, ``{ASTRA-5G: Automated Over-the-Air Security Testing and Research Architecture for 5G SA Devices},'' in \emph{ACM WiSec}, 2024.

\bibitem{rrc}
3GPP, ``Ts 38.331, v18.5.0,'' 2024.

\bibitem{nas}
3GPP, ``Ts 24.501, v18.10.0,'' 2025.

\end{thebibliography}

\end{document}